\documentclass[journal]{IEEEtran}

\usepackage{cite}

\ifCLASSINFOpdf
   \usepackage[pdftex]{graphicx}
   \graphicspath{{../pdf/}{../jpeg/}}
   \DeclareGraphicsExtensions{.pdf,.jpeg,.png}
\else
\fi
\UseRawInputEncoding
\usepackage{amsmath}
\usepackage{array}
\usepackage{standalone}
\usepackage{etoolbox}
\newtoggle{arXiv}
\newtoggle{bibrun}
\usepackage{newtxtext, newtxmath}
\usepackage{graphicx}
\usepackage{color, calc}
\usepackage{array, booktabs}
\usepackage{tikz, tikz-3dplot}
\usetikzlibrary{arrows, arrows.meta, calc, positioning, decorations.pathmorphing}
\tikzset{>=Stealth}
\usepackage{siunitx}

\usepackage{enumitem}
\setlist[description]{labelindent=0pt, leftmargin=\parindent, font=\normalfont\itshape}
\usepackage{url}
\usepackage{subfigure}
\usepackage{textcomp}
\usepackage{eurosym}
\usepackage{xfrac}
\usepackage{pgfplots}
\usepackage{stmaryrd}
\pgfplotsset{compat=1.17}
\usepackage{makecell}
\usepackage[utf8]{inputenc}
\usepackage{hyperref}

\newcommand{\mg}{MaRGE}

\begin{document}

\title{A fully open-source framework for streaming and cloud-processing of low-field MRI data}

\author{\IEEEauthorblockN{
Teresa~Guallart-Naval\IEEEauthorrefmark{1},
John~Stairs\IEEEauthorrefmark{2},
Jos\'e~M.~Algar\'in\IEEEauthorrefmark{1}$^,$\IEEEauthorrefmark{6},
Hui~Xue\IEEEauthorrefmark{2},
Juan~Benlloch\IEEEauthorrefmark{6},
Pablo~Benlloch\IEEEauthorrefmark{1}$^,$\IEEEauthorrefmark{6},
Jose~Borreguero\IEEEauthorrefmark{1},
Jesús~Conejero\IEEEauthorrefmark{1},
Fernando~Galve\IEEEauthorrefmark{1},
Pablo~García-Cristóbal\IEEEauthorrefmark{1},
Miguel~Lacalle\IEEEauthorrefmark{6},
Beatrice~Lena\IEEEauthorrefmark{4},
Laia~Porcar\IEEEauthorrefmark{1}$^,$\IEEEauthorrefmark{3}
Steven~J.~Schiff\IEEEauthorrefmark{5},
Andrew~Webb\IEEEauthorrefmark{4},
Michael~Hansen\IEEEauthorrefmark{2},
and~Joseba~Alonso\IEEEauthorrefmark{1}}

\IEEEauthorblockA{\IEEEauthorrefmark{1}MRILab, Institute for Molecular Imaging and Instrumentation (i3M), Consejo Superior de Investigaciones Cient\'ificas (CSIC) \& Universitat Polit\`ecnica de Val\`encia (UPV), Valencia, Spain}\\
\IEEEauthorblockA{\IEEEauthorrefmark{2}Microsoft Research, Health Futures, Washington, USA}\\
\IEEEauthorblockA{\IEEEauthorrefmark{6}Full Body Insight. S.L., Paterna, Spain}\\
\IEEEauthorblockA{\IEEEauthorrefmark{3}PhysioMRI Tech. S.L., Paterna, Spain}\\
\IEEEauthorblockA{\IEEEauthorrefmark{4}Leiden University Medical Center (LUMC), Dept. of Radiology, Leiden, Netherlands}\\
\IEEEauthorblockA{\IEEEauthorrefmark{5}Yale University, Dept. of Neurosurgery and Dept. of Epidemiology of Microbial Diseases, New Haven, USA}

\thanks{Corresponding author: T. Guallart-Naval (tguanav@i3m.upv.es).}
}


\maketitle

\begin{abstract}
\newline
Purpose: {\normalfont To present a fully open-source framework for quasi-real-time streaming and cloud-based processing of low-field (LF) MRI data, addressing the growing computational demands of advanced reconstruction and post-processing pipelines in portable and affordable MRI systems.}\\
Methods: {\normalfont The proposed framework integrates open-source scanner control software with a network-enabled streaming architecture, allowing for raw data to be transmitted directly from the MRI console to remote compute resources. Cloud-based processing modules support image reconstruction and advanced post-processing, including computationally intensive physics- and learning-based methods, while maintaining compatibility with low-cost on-device control hardware.}\\
Results: {\normalfont The system enables continuous acquisition-to-reconstruction workflows in LF-MRI without requiring specialized high-performance console architectures. Selected example applications include deep-learning-based denoising, field-induced distortion correction, and non-Cartesian image reconstruction. Experimental demonstrations show reliable streaming performance.}\\
Conclusions: {\normalfont Open-source streaming and cloud-processing provide an effective pathway to overcome the computational limitations of embedded LF-MRI consoles. By decoupling acquisition hardware from intensive reconstruction workloads, the proposed framework supports scalable deployment of advanced algorithms while preserving the affordability and portability that motivate LF-MRI.}
\end{abstract}

\IEEEpeerreviewmaketitle

\section{Introduction}

\IEEEPARstart{M}{agnetic} resonance imaging (MRI) systems are fundamentally orchestrated by their control computers, which serve as the central hub for user interaction, data handling, and image reconstruction. In low-field MRI (LF-MRI), this computational layer is becoming increasingly critical as the field transitions from minimal viable hardware toward clinically useful performance and image quality through advanced processing.

Portable and affordable LF-MRI systems have developed rapidly in recent years, enabled in large part by permanent-magnet architectures that avoid superconducting infrastructure. By substantially reducing cost, installation requirements, and operational complexity, these scanners can be deployed in point-of-care \cite{McDaniel2019,Nakagomi2019,Cooley2020,OReilly2020,Sheth2021,Mazurek2021,Liu2021}, home-care \cite{Guallart-Naval2022}, or resource-limited settings \cite{Obungoloch2023,Guallart-Naval2025c}. However, operation at low magnetic field strengths entails intrinsic constraints, most notably reduced signal-to-noise ratio (SNR) \cite{Sarracanie2020,Webb2023b}, which translates into noisier images, reduced spatial resolution, and longer acquisition times. Furthermore, many portable LF systems rely on compact permanent magnet designs, which exhibit pronounced magnetic field ($B_0$) inhomogeneities, leading to signal degradation and geometric distortions \cite{Borreguero2025b}.

To mitigate these limitations, a wide range of advanced reconstruction and post-processing techniques have been proposed, including deep learning-based pipelines \cite{Zhao2024,Islam2023,Shimron2025} and physics-based model-driven approaches \cite{Koolstra2021,Borreguero2025}. While effective, many of these techniques are computationally demanding and typically require substantial resources. In contrast, the control computers integrated into affordable and portable low-field MRI systems are often designed under strict power, size, and cost constraints, making on-premise execution of such methods impractical. Dedicated high-performance console architectures have recently been proposed to address this limitation \cite{Schote2025}. However, such solutions rely on specialized hardware and increase overall system complexity and cost.

\begin{figure*}
  \centering
  \includegraphics[width=0.85\textwidth]{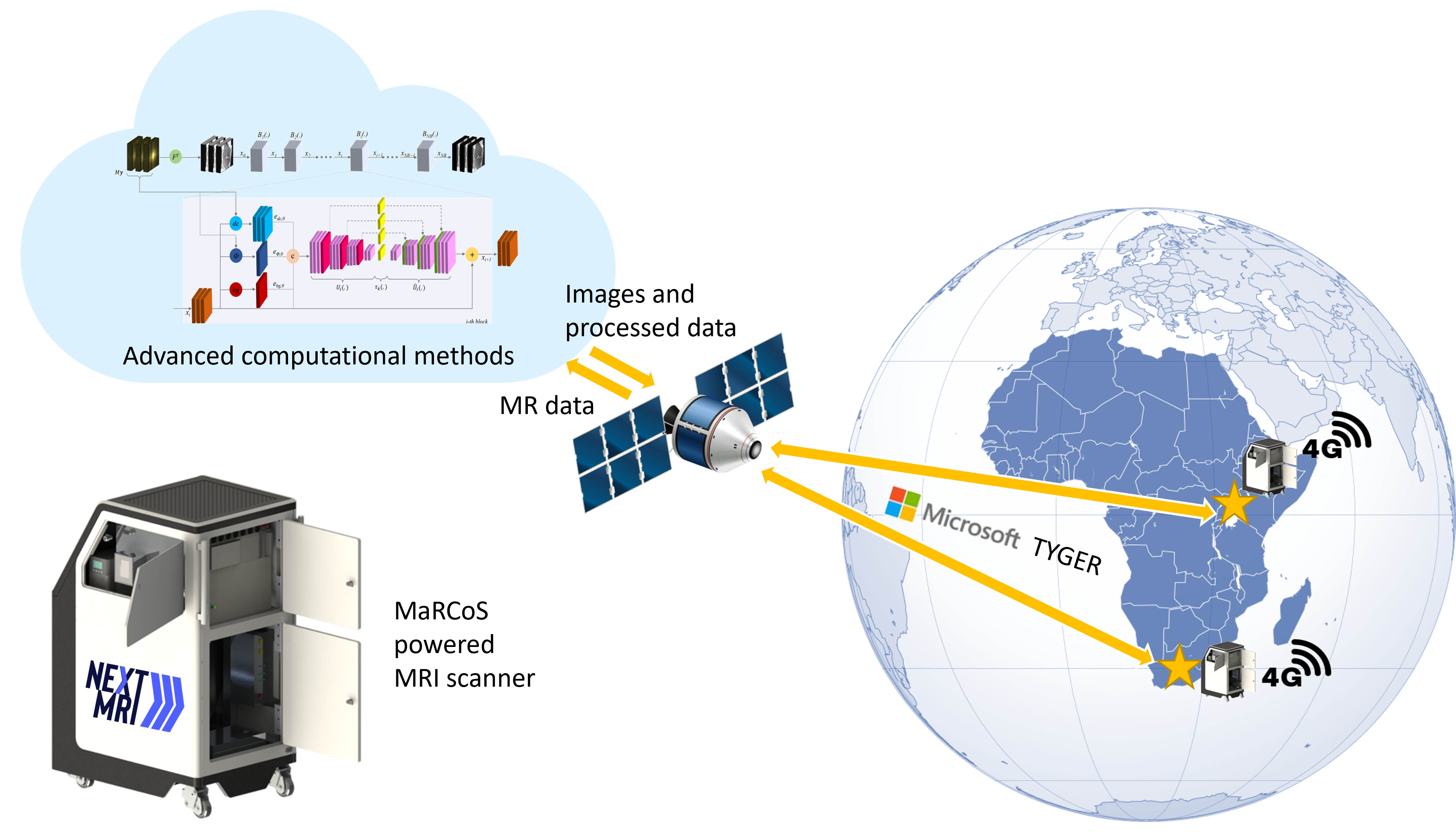}
  \caption{Scheme for cloud processing of streamed data from
MaRCoS-powered MRI scanners, including portable and point-of-care
systems. Tyger enables this connection, which can be accessed
globally through a mobile network.}
  \label{fig:squeme}
\end{figure*}

In this work, we address this bottleneck by inline-integrating cloud-based data processing directly into the low-field MRI acquisition workflow. Specifically, we interface the open-source MaRCoS control system \cite{Negnevitsky2023,Guallart-Naval2022b} and its most sophisticated graphical environment (MaRGE, \cite{Algarin2024}) with Tyger, an open-source platform for remote signal processing \cite{Tyger}. We thereby present the first fully open-source framework for cloud data processing with a specific focus on LF-MRI, as illustrated in Fig.~\ref{fig:squeme}. The integration is validated using a portable elliptical Halbach scanner (NextMRI, \cite{Galve2024}) and executing algorithms for distortion correction, non-Cartesian $k$-space reconstructions, and deep learning techniques. With this development, affordable MRI systems deployed anywhere in the world can leverage powerful cloud computing resources for data processing and image reconstruction. Additionally, we show that this connectivity can be achieved using standard mobile network infrastructures.

\section{Technical background}

This section introduces the software tools integrated in this work, together with the low-field MRI scanner used for experimental evaluation.

\subsection{MaRCoS / MaRGE: control system and GUI.}

MaRCoS (Magnetic Resonance Control System) is an open-source control system designed to enable high-performance MRI experiments using low-cost and accessible hardware platforms, with a particular focus on LF systems \cite{Negnevitsky2023,Guallart-Naval2022b}. Its architecture has been conceived for flexibility, providing full control over radio-frequency (RF) transmission and reception, gradient waveforms, and data acquisition.

MaRGE is a graphical user interface (GUI) we have built for MaRCoS to facilitate scanner operation while retaining access to advanced acquisition and reconstruction options \cite{Algarin2024}. Implemented in Python, MaRGE integrates vendor-neutral sequence design and execution (compatible with Pypulseq \cite{Ravi2019}), data handling, and standard Fourier reconstruction within the acquisition interface using the computational resources available on the scanner control computer, while also providing a separate post-processing environment for the analysis and processing of previously acquired raw data.

\subsection{Tyger: remote signal processing tool.}

Tyger is an open-source platform for remote signal processing, designed to enable the execution of computational pipelines outside the data acquisition site \cite{Tyger}. It allows data generated by distributed systems to be streamed to remote computing resources, including cloud infrastructures, where processing can be performed independently of the hardware available at the source.

A central design principle is its language- and framework-independent approach to signal processing, allowing existing processes to run without modification \cite{Hansen2013, BART} while also supporting the development of custom processing routines. Processing code can be written in any programming language, provided it supports simple byte stream-based input and output, implemented through named pipes for data transfer. Tyger does not rely on a dedicated software development kit (SDK), enabling processes to be developed and tested locally and then packaged into a Docker container. The same container can be executed remotely, with the run configured through a file written in a lightweight human-friendly data-serialization language file (\texttt{.yaml}) specifying the Docker image and the processing command to be executed. Tyger supports deployment on both public cloud infrastructures (e.g., Microsoft Azure) and user-managed remote computing resources, such as privately hosted GPU servers.

\subsection{NextMRI: low-field MRI scanner.}
\label{subsec:nextmri}

The NextMRI system is a portable low-field MRI scanner designed for extremity and brain imaging \cite{Galve2024} and located at the MRILab in Valencia, Spain, during these experiments. The system is based on a 44\,cm-long  elliptical permanent magnet in a Halbach configuration. The magnet provides a field of $\sim$90\,mT with an homogeneity of approximately 5,000\,ppm over a 20\,cm diameter spherical volume. The scanner is equipped with full-length gradient coils manufactured from water-jetted copper plates shaped with elliptical molds, resulting in low electrical resistance and enabling high-current operation without excessive heating. The gradient system provides efficiencies of $(0.61,\,0.87,\,0.87)$\,mT/m/A along the three axes and is driven by three independent gradient power amplifiers (AE Techron~7224), capable of delivering currents up to 50\,A per channel. The RF subsystem employs interchangeable transmit-receive coils: a solenoidal knee coil (15\,cm diameter, 15\,cm length), a head coil based on a helmet-like geometry~\cite{Sarracanie2020}, and a larger elliptical coil (26\,cm length, 19.8/26.8\,cm on the short/long axis). The complete system is self-contained, mounted on a mobile platform, and operates from a standard single-phase power outlet. The system layout follows previously reported cable routing and shielding guidelines~\cite{GuallartNaval2026a} to enable operation close to the thermal noise limit. The scanner is controlled by an industrial PC (TBOX-2825) equipped with an Intel Core i5-8265U processor and 16\,GB RAM.


\section{Methods}

This section describes the technical details of the integration of Tyger into the MaRCoS/MaRGE ecosystem, together with data transfer measurements to benchmark against real-world operating conditions and some selected processing workflows showcasing potentially relevant low-field MRI applications. In addition, we describe the experimental measurements performed to validate the proposed framework. All experiments presented in this section are conducted by connecting to Microsoft Azure through the Tyger Technology Evaluation Program (Tyger TEP), using an NVIDIA A100 80\,GB GPU in a server located in France.

\subsection{Data transfer measurements}

\begin{figure}
  \centering
  \includegraphics[width=0.85\columnwidth]{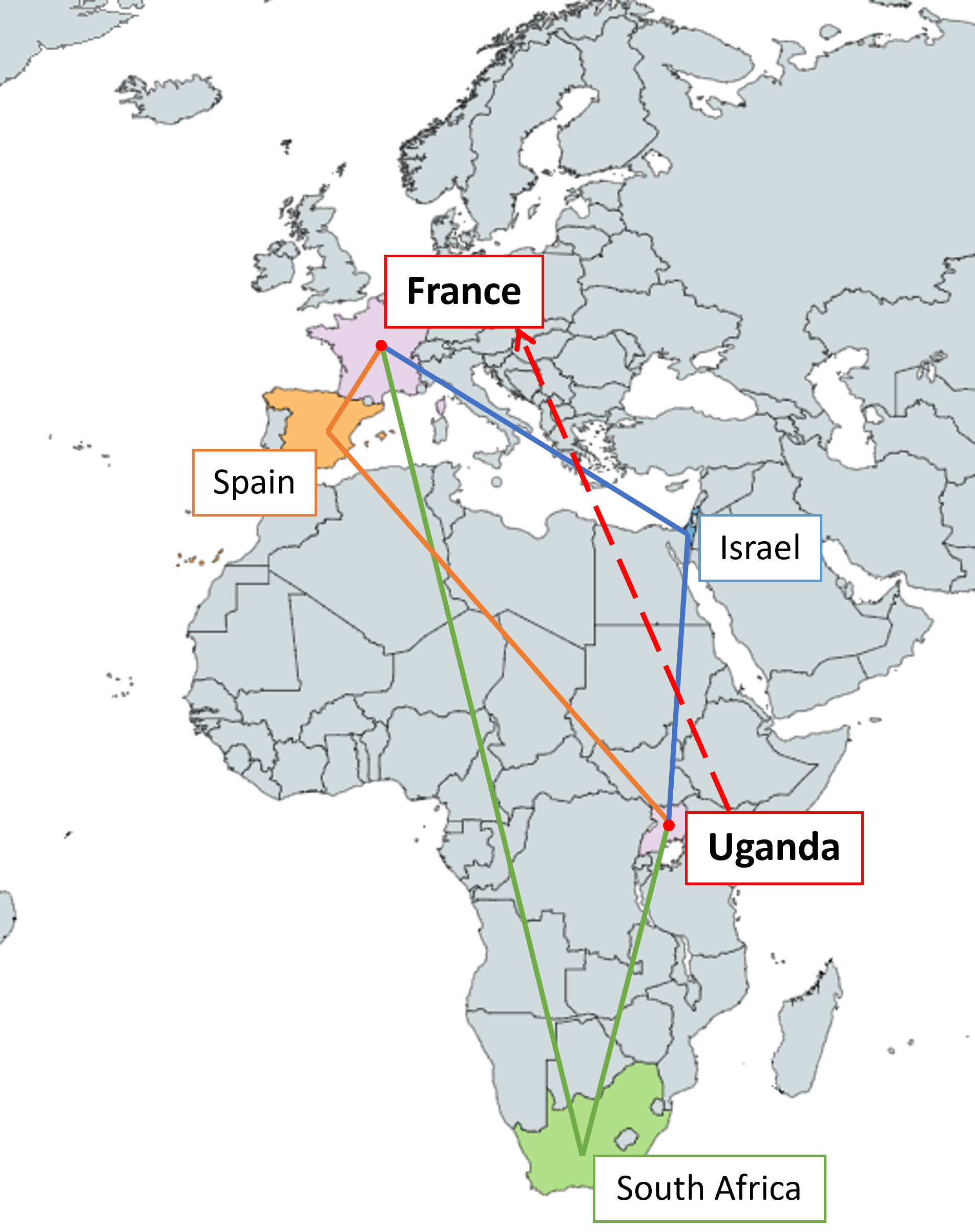}
  \caption{Network routing configurations tested for the data trransfer measurements. Data were transmitted from Uganda to France using different paths over the underlying cloud infrastructure, routed via Spain, Israel, and South Africa.}
  \label{fig:latencies}
\end{figure}

To evaluate the feasibility of remote data processing under real-world network conditions, we measured data transfer times from Uganda, selected as a representative low-income setting with heterogeneous connectivity. Datasets of 32\,MB, typical of a low-field brain acquisition, were transmitted to a cloud computing node in France using Tyger. These measurements were conducted using a Wi-Fi network available at MUST (Mbarara University of Science and Technology), as well as 4G and 3G mobile connections, and through multiple available network paths over the underlying cloud infrastructure. The tested routing configurations are depicted in Fig.~\ref{fig:latencies}. For reference, an additional measurement was tested from Spain (selected as a representative high-income setting) using a standard household Wi-Fi connection. For each test, the upload time was recorded. All measurements in this subsection are performed outside the MaRGE environment using a standard computer rather than the MRI scanner.

\subsection{MaRGE-Tyger integration}

\begin{figure*}
  \centering
  \includegraphics[width=0.85\textwidth]{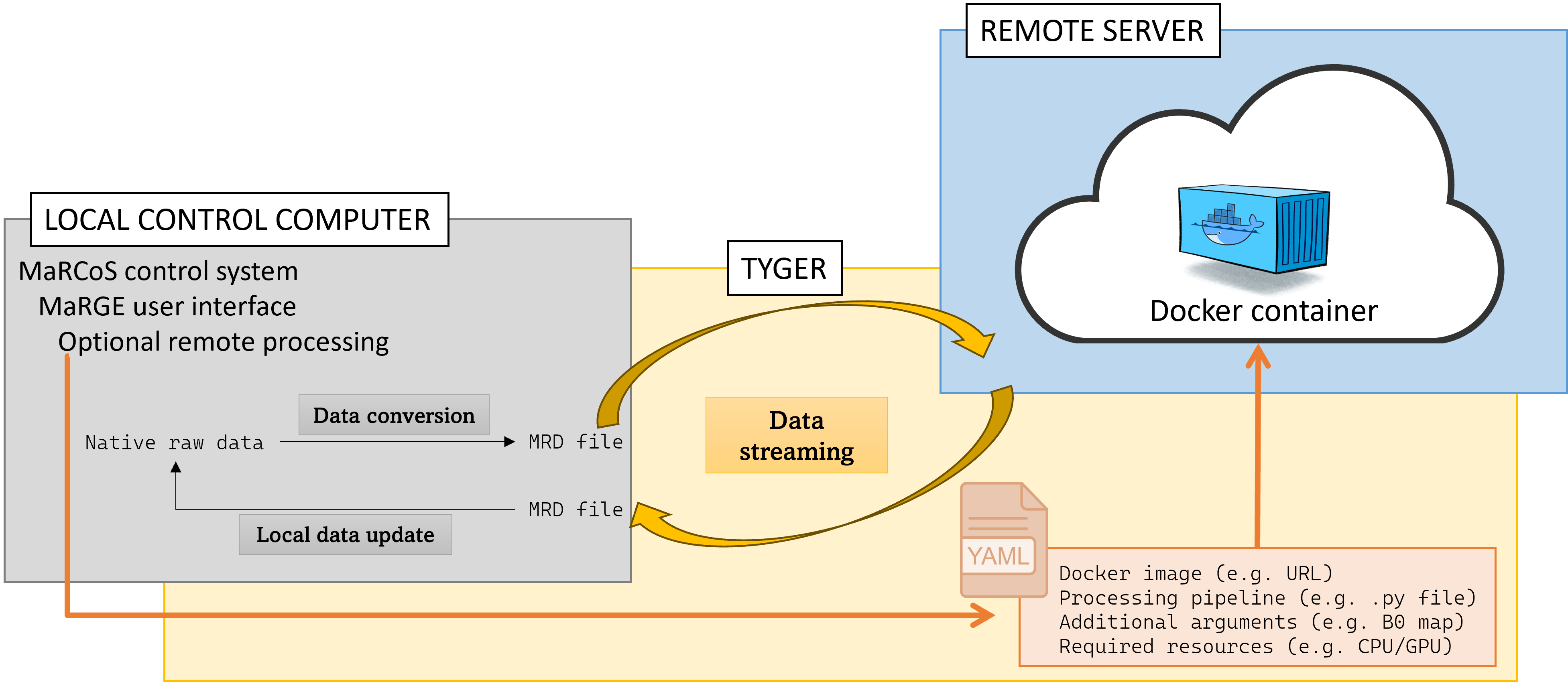}
  \caption{Architecture of the MaRGE-Tyger integration. Native MaRGE raw data (\texttt{.mat}) are converted locally into the open MRD format and accompanied by a YAML configuration file specifying the selected processing pipeline and arguments. Data are streamed through Tyger to a remote computing node, where a Docker container is executed. Reconstructed data are streamed back to MaRGE for visualization.}
  \label{fig:integration}
\end{figure*}

Figure~\ref{fig:integration} illustrates the architecture of the MaRGE-Tyger integration. Native MaRGE raw data files are converted locally from the \texttt{.mat} format into the open MRD standard \cite{MRD}. Additionally, MaRGE generates a \texttt{.yaml} configuration file specifying the selected processing pipeline and its required inputs, which are defined directly within the graphical interface. The MRD dataset and configuration file are streamed to a remote computing resource through Tyger, where the requested pipeline is executed within a Docker container. When operating in a cloud environment (e.g., Azure), the container image is retrieved from a public registry; alternatively, it can be pre-deployed on a remote node. After execution, the reconstructed data are streamed back and handled directly within the MaRGE interface.

This integration enables automatic remote processing as part of the standard scanning workflow. Authenticated access to Tyger services is provided through a login button in the graphical interface (Fig.~\ref{fig:phantom}a, label~1). Selection of the reconstruction or processing pipeline is performed through the dedicated Tyger tab (Fig.~\ref{fig:phantom}a, label~2). Reconstructed images are returned and displayed as part of the acquisition output without requiring additional user intervention beyond sequence execution. To avoid interrupting scanner operation, MaRGE supports parallel execution, allowing new sequences to be launched while previous datasets are being transmitted or reconstructed remotely.

Tyger functionality is also available in the post-processing interface, where a dedicated Tyger tab is provided (Fig.~\ref{fig:phantom}b, label~3). This interface allows users to apply remote pipelines to previously acquired raw datasets. All features described in this section are implemented in the current version of the MaRGE repository (see \emph{Code availability}).

\subsection{Selected compute-intensive applications}
To illustrate the capabilities of the proposed integration, we attempted to address some key limitations characteristic of LF-MRI systems. The following pipelines can be enabled through the Tyger tab (labels 2\&3 in Fig.~\ref{fig:phantom}) within MaRGE for the corresponding sequences.

\subsubsection{Denoising} 
We implemented a deep-learning-based denoising pipeline using a pre-trained model obtained with the SNRAware method \cite{Xue2025}, provided by Microsoft Research through Tyger TEP. The model was trained exclusively on high-field data. Denoising was implemented for RARE (Rapid Acquisition with Relaxation Enhancement, \cite{Hennig1986}) acquisitions within the MaRGE workflow. To ensure correct operation of the model, input images are scaled to SNR units \cite{Kellman2005}. For this purpose, additional noise-only readouts are acquired as part of the RARE sequence.

\subsubsection{Distortion correction}
We corrected geometric distortions in RARE acquisitions using conjugate-phase (CP) reconstruction \cite{Noll1991,Noll2005}. CP compensates geometric distortions due to  $B_0$ field inhomogeneities by explicitly accounting for the phase evolution induced by off-resonance effects during signal readout. Rather than assuming a uniform frequency encoding, the reconstruction incorporates prior knowledge of the spatially varying $B_0$ field to correct the effective sampling positions along the readout direction. Although non-iterative, this approach requires explicit evaluation of a voxel-dependent phase term across the reconstructed volume and for multiple acquisition time points. As a result, the CP correction is computationally more demanding than a conventional Fourier-based reconstruction and is unsuitable for execution on the scanner control computer.

The $B_0$ map is obtained using the Single-Point Double-Shot (SPDS) method \cite{Borreguero2025}, which is integrated as a dedicated sequence within MaRGE. Following SPDS acquisition, the measured field map is automatically fitted using a polynomial model, and the resulting fit is stored in a text file. This file can then be selected within MaRGE when distortion correction through Tyger is enabled.

Distortion correction can be combined with denoising by enabling simultaneously both options in MaRGE. Denoising takes place first, as it relies on normalization of the input data to SNR units, which cannot be accurately applied after distortion correction due to the modification of the noise background introduced by the transformation.

\subsubsection{Non-Cartesian reconstructions}
Finally, we evaluate the use of remote processing for non-Cartesian reconstructions from PETRA acquisitions (Pointwise Encoding Time Reduction with Radial Acquisition, \cite{Grodzki2012}), where $k$-space is sampled along radial trajectories and cannot be reconstructed using conventional Fourier-based methods. Instead, reconstructions were performed using iterative Algebraic Reconstruction Techniques (ART) \cite{Gordon1970}, where the image is generated by estimating the solution to a linear system defined by the non-Cartesian encoding operator and the measured $k$-space samples \cite{Algarin2020}. This formulation requires repeated forward and adjoint evaluations of the encoding model and is impractical to execute on a regular control computer.

\subsection{Phantom validation}

\begin{table*}
\caption{Acquisition parameters for the RARE experiments used in phantom and \emph{in-vivo} studies. Reported reconstruction times correspond to total cloud execution time, including data transfer, processing, and return of reconstructed data. When applicable, local reconstruction times are also reported for comparison. PF refers to Partial Fourier undersampling along the second phase-encoding direction.}
\centering
\resizebox{\textwidth}{!}{%
\begin{tabular}{c c c c c c c c c c c c c c c}
\toprule
Figure & \thead{FoV (mm$^3$)} & \# pixels &  \thead{PF  (\%)} & \thead{Res. (mm$^3$)} & \thead{BW \\ (kHz)} & \thead{TR \\(ms)} & \thead{TE \\(ms)} & \thead{ETL} & \thead{TI \\(ms)} & \thead{$k$-space \\filling} & Avgs. & \thead{Scan time \\(min)} & \thead{Tyger recon. \\ time (s)} &\thead{Local recon. \\ time} \\

\midrule
\ref{fig:phantom}-a) & $120\times120\times120$ & $128\times 128\times 28$  & 80 & $0.9\times0.9\times4.3$ & 30 & 500 & 20 & 5 & -- & center-out  & 1 & 4.4 & 21 & 1 h 25 min\\
\midrule

\ref{fig:phantom}-b) & $150\times150\times150$ & $320\times 320\times 36$  & 100 & $0.5\times0.5\times4.2$ & 27 & 200 & 20 & 5 & -- & center-out  & 1 & 7.7 & 91 & --\\
\midrule

\ref{fig:knee_den} & $160\times160\times160$ & $500\times 500\times 50$  & 100 & $0.3\times0.3\times3.2$ & 42 & 200 & 20 & 5 & -- & center-out  & 1 & 16.7 & 266 & --\\
\midrule

\ref{fig:brain}-a)& $208\times224\times19$ & $130\times 136\times 38$  & 70 & $1.6\times1.6\times5$ & 43 & 2500 & 40 & 8 & -- & linear  & 1 & 18.8 & 84 & --\\
\midrule

\ref{fig:brain}-b) & $208\times224\times19$ & $130\times 136\times 38$  & 70 & $1.6\times1.6\times5$ & 43 & 1200 & 40 & 4 & 120 & center-out  & 1 & 18 & 80 & --\\
\bottomrule

\end{tabular}}
\label{tab:RARE_params}
\end{table*}

\begin{table*}
\caption{Acquisition parameters for the SPDS field mapping experiments used for distortion correction.}
\centering
\resizebox{\textwidth}{!}{%
\begin{tabular}{c c c c c c c c c c c c}
\toprule
Figure & \thead{FoV \\ (mm$^3$)} & \# pixels &  \thead{Flip \\ angle (º)} & \thead{RF pulse \\ time (us)} & \thead{TR \\ (ms)} & \thead{Dummy \\ pulses} & \thead{Dead \\ times (us)} & \thead{Avgs.} & \thead{Scan time \\(min)} & \thead{Zero padding \\ order} &  \thead{Polynomial \\ fit order}  \\

\midrule
\ref{fig:phantom}-a) & $160\times160\times160$ & $30\times 30\times 30$  & 40 & 36 & 20 & 10 & 320 / 380 & 1 & 9.6 & 4 & 8 \\
\midrule

\ref{fig:brain} & $240\times240\times240$ & $30\times 30\times 30$  & 80 & 20 & 20 & 10 & 320 / 410 & 1 & 9.6 & 4 & 5 \\
\bottomrule
\end{tabular}}
\label{tab:SPDS_params}
\end{table*}

\begin{table*}
\caption{Acquisition parameters for the PETRA experiments. Reconstruction times are reported for cloud execution through Tyger and for local reconstruction performed offline as a reference.}
\centering
\resizebox{\textwidth}{!}{%
\begin{tabular}{c c c c c c c c c c c c c c}
\toprule
Figure & \thead{FoV (mm$^3$)} & \thead{Flip \\ angle (º)} &  \thead{RF pulse \\ time (us)} & \thead{Res. (mm$^3$)} & \thead{BW \\ (kHz)} & \thead{TR \\(ms)} & \thead{Radial spokes/ \\ pointwise} & \thead{US \\ rad} & \thead{Dead time (us)/ \\ acq time (ms)} & Avgs. & \thead{Scan time \\(min)} & \thead{Tyger recon. \\ time (s)} &\thead{Local recon. \\ time} \\

\midrule
\ref{fig:petra} & $200\times160\times160$ & 60  & 30 & $1.7\times1.6\times1.6$ & 48 & 20 & 9488 / 10752 & 4 & 300/1.25 & 2 & 13.5 & 204 & 19 h 49 min \\
\bottomrule
\end{tabular}}
\label{tab:PETRA_params}
\end{table*}

Phantom experiments were performed to validate the integration and to test the execution of application-specific pipelines through both the scanner-control and post-processing graphical interfaces. To this end, we imaged a commercial phantom \cite{ACRphantom}, which provides a known reference structure, using the small RF coil described in Section~\ref{subsec:nextmri}. Specifically, we acquired two RARE datasets, which are used to apply the distortion-correction and denoising procedures. For distortion correction, we mapped the $B_0$ field with an homogeneous phantom filled with a copper sulfate solution using the whole space inside the RF coil.

The acquisition parameters for the RARE and SPDS sequences are summarized in Tables~\ref{tab:RARE_params} and~\ref{tab:SPDS_params}, respectively. For all phantom experiments, we measured the total cloud reconstruction time, including data transfer, processing, and return of the reconstructed data. In the case of distortion correction, we also performed a local reconstruction as a reference for comparison with cloud execution.

\subsection{\emph{In-vivo} experiments}

We performed \emph{in-vivo} experiments to demonstrate representative use cases of the proposed integration. In all cases, a Faraday blanket was used to reduce external electromagnetic interference \cite{GuallartNaval2026a} and, as in the phantom acquisitions, total cloud reconstruction times were measured. All RF coils employed in this section are described in Section~\ref{subsec:nextmri}. Acquisition parameters for the RARE, SPDS, and PETRA sequences are summarized in Tables~\ref{tab:RARE_params},~\ref{tab:SPDS_params}, and~\ref{tab:PETRA_params}, respectively. Each \emph{in-vivo} image shown corresponds to a different volunteer.

Knee imaging was performed using the small RF coil. On the one hand, a sagittal T1-weighted (T1-w) RARE acquisition with an in-plane resolution of $0.3\times0.3$\,mm$^2$ was acquired and processed using the denoising pipeline. On the other hand, a PETRA acquisition was performed to validate non-Cartesian reconstruction. For this dataset, we also performed the corresponding local reconstruction offline as a reference.

Brain images were acquired with the helmet coil using transverse RARE acquisitions with T2-w and inversion-recovery (IR) contrast. These datasets were first processed using the denoising pipeline and subsequently corrected for geometric distortions. SPDS field maps were acquired with the large elliptical RF coil and an homogeneous water and copper sulfate phantom, providing full coverage of the brain field of view.

\section{Results}

\begin{table}
\caption{Data transfer time (in s) measured for remote execution through Tyger using 32~MB datasets, representative of a typical low-field brain acquisition. Measurements were performed from Uganda using different network paths and connection types, and compared with a reference measurement acquired from Spain using a standard household Wi-Fi connection. Reported values include data transmission to the remote computing resource.}
\centering
\begin{tabular}{l l c c c}
\toprule
 &  & South Africa & Israel & Spain \\
\midrule
 Uganda & WiFi & 7 & 8 & 8 \\
 & 4G   & 19 & 19 & 16 \\
 & 3G   & 76 & 107 & -- \\
\midrule
Spain & WiFi & 2 & 2 & 1 \\
\bottomrule
\end{tabular}
\label{tab:latencies}
\end{table}

Table~\ref{tab:latencies} presents the data transfer times measured for remote execution through Tyger under different network paths and connection types. Measurements were successfully completed for all configurations, except for the 3G connection from Uganda routed through Spain, which failed in repeated attempts.

\begin{figure*}
  \centering
  \includegraphics[width=0.8\textwidth]{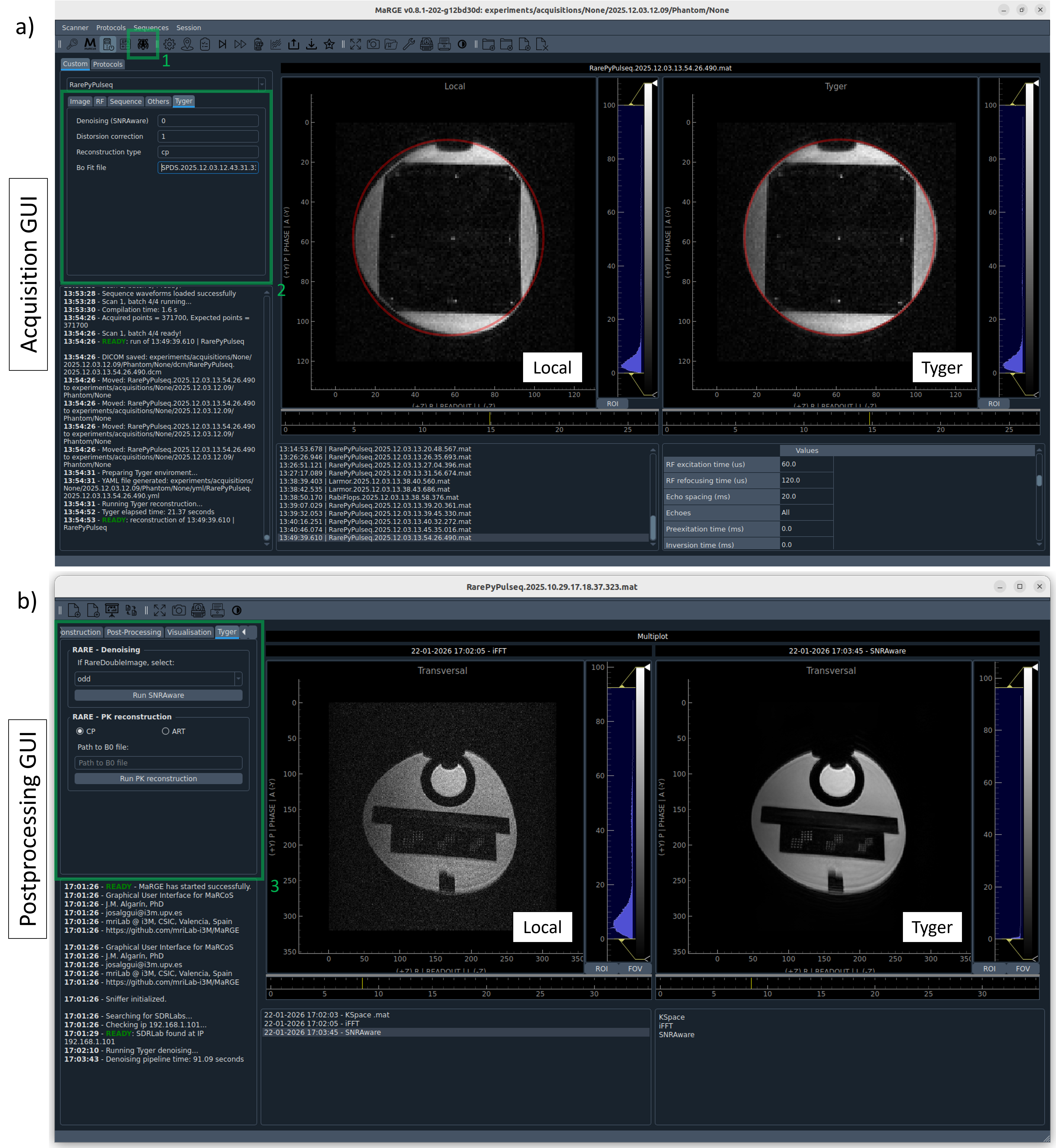}
   \caption{Phantom experiments used to validate the integration of Tyger within the MaRGE. In both panels, the standard Fourier reconstruction is shown alongside the corresponding cloud-processed result obtained via Tyger. (a) MaRGE acquisition GUI for a RARE phantom acquisition with distortion correction applied through Tyger. Label~1 indicates the Tyger connection button, and label~2 the reconstruction selection panel within the RARE sequence. Red contours indicate the expected phantom geometry. (b) MaRGE post-processing interface for a different phantom acquisition with denoising applied through Tyger. Label~3 highlights the dedicated Tyger panel in the post-processing window.}
  \label{fig:phantom}
\end{figure*}

Figure~\ref{fig:phantom} shows phantom experiments used to validate the integration of Tyger within MaRGE. Figure~\ref{fig:phantom}a displays the acquisition GUI for a RARE phantom experiment, where distortion correction is applied through Tyger. Figure~\ref{fig:phantom}b shows the post-processing GUI for a different phantom acquisition processed using the denoising pipeline. In both views, the local reconstruction is shown on the left, with the corresponding cloud-processed result on the right. Reconstruction logs displayed in the interface report the execution times for cloud processing, which are summarized in Table~\ref{tab:RARE_params}.

\begin{figure*}
  \centering
  \includegraphics[width=\textwidth]{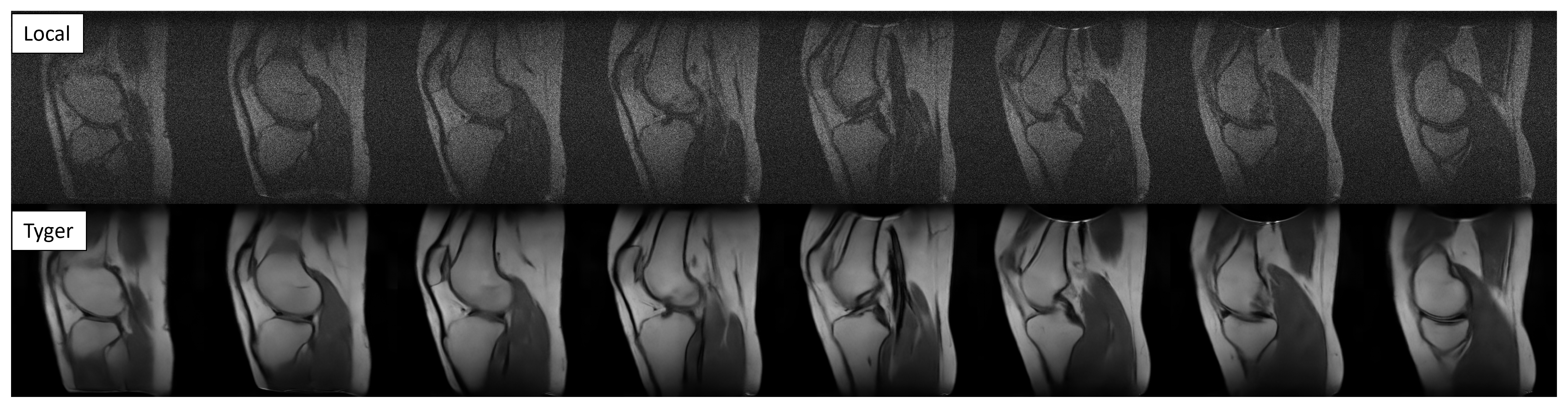}
  \caption{\emph{In-vivo} knee acquisition showing the locally reconstructed image (top) and the corresponding Tyger result after denoising (bottom).}
  \label{fig:knee_den}
\end{figure*}

Figure~\ref{fig:knee_den} shows a knee acquisition, comparing the image reconstructed locally on the scanner computer with the corresponding result obtained through Tyger after denoising.

\begin{figure}
  \centering
  \includegraphics[width=\columnwidth]{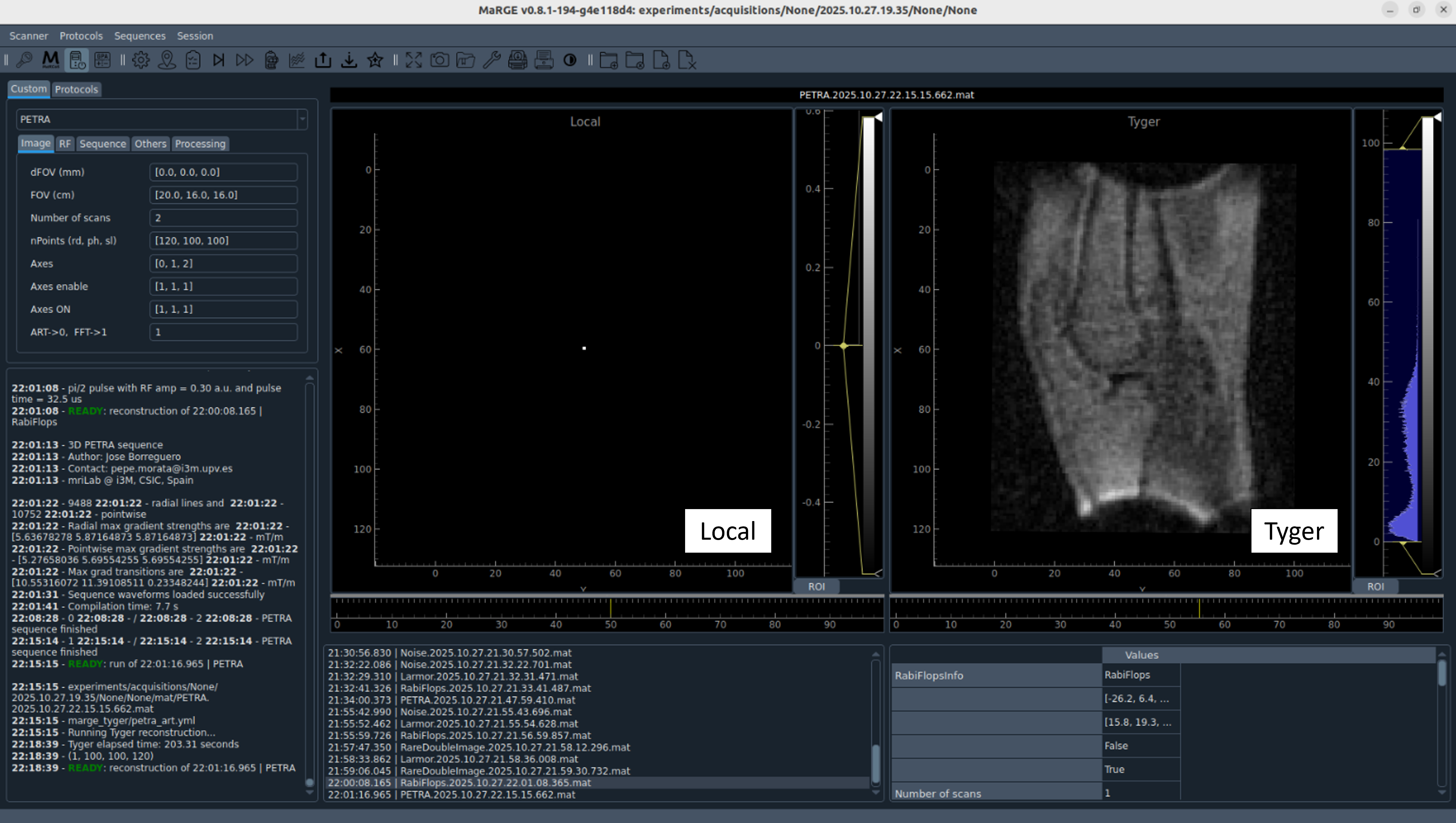}
  \caption{Knee PETRA acquisition displayed in the acquisition GUI. No local reconstruction is available during acquisition, resulting in an empty local reconstruction panel, while the corresponding Tyger-based reconstruction is shown on the right.}
  \label{fig:petra}
\end{figure}

Figure~\ref{fig:petra} displays a knee PETRA dataset in the acquisition GUI. During acquisition, no image is available from the local reconstruction, resulting in an empty panel, while the Tyger-based reconstruction is displayed. The remote and local reconstruction times measured offline are reported in Table~\ref{tab:PETRA_params}.

\begin{figure*}
  \centering
  \includegraphics[width=\textwidth]{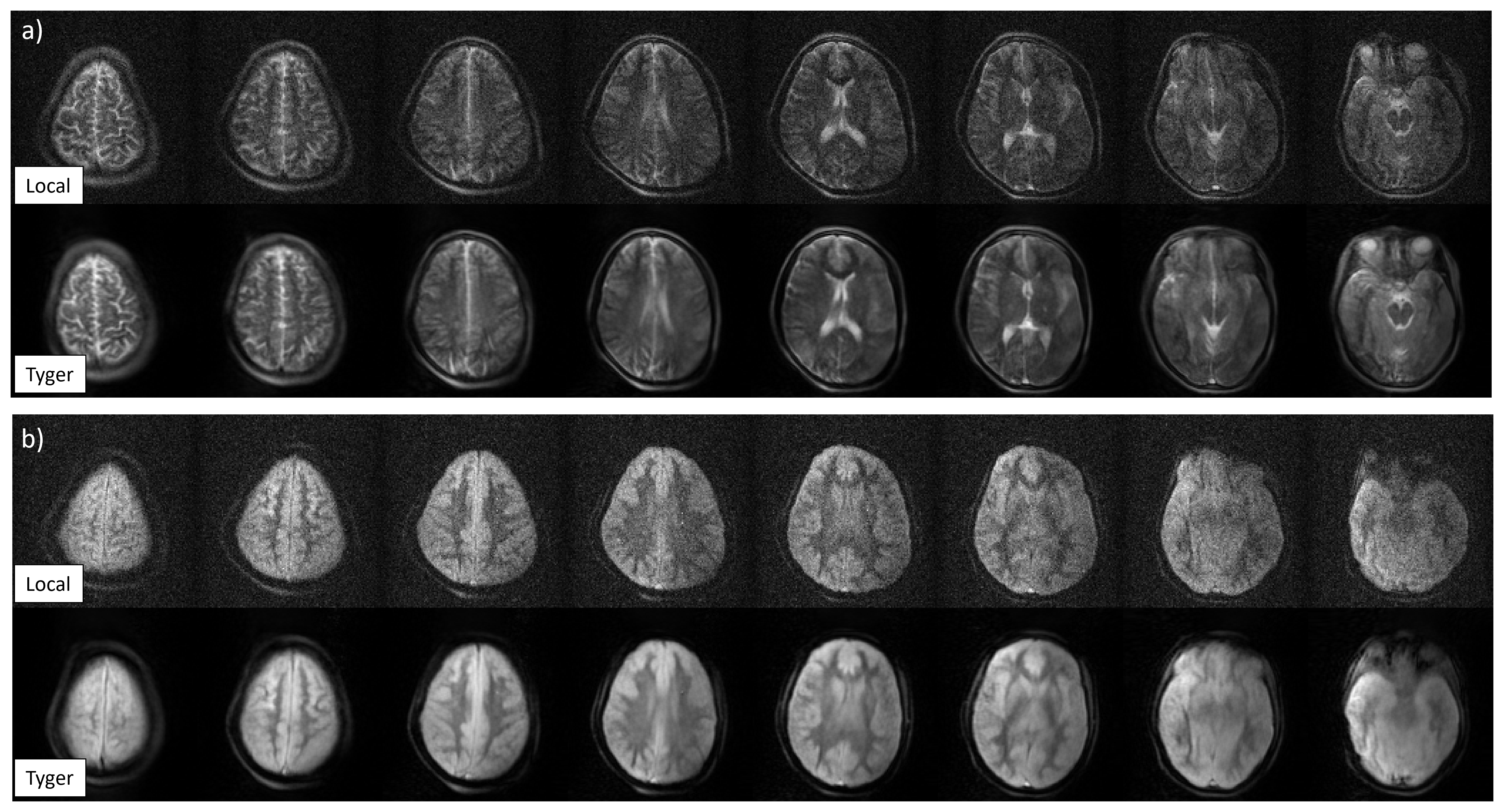}
  \caption{\emph{In-vivo} brain acquisitions from two different volunteers: (a) T2-weighted and (b) inversion recovery. For each case, the local reconstruction is shown on top, and the Tyger-processed image with denoising and distortion correction is shown below.}
  \label{fig:brain}
\end{figure*}

Figure~\ref{fig:brain} shows brain images acquired from two different volunteers using two different contrasts: (a) T2-weighted and (b) inversion recovery. For each case, selected slices are shown along the second phase-encoding direction, comparing locally reconstructed images with the corresponding Tyger results obtained after denoising and distortion correction.

\section{Discussion}

The data transfer measurements demonstrate that cloud-based MRI reconstruction can be executed reliably from geographically remote locations, including low-income settings with limited network infrastructure (Table~\ref{tab:latencies}). Successful data transfer across heterogeneous network conditions shows that the proposed framework is not restricted to well-connected environments. Although transfer times increased under lower-bandwidth connections, particularly with 3G mobile networks, cloud-based processing remained feasible in all tested scenarios.

Figure~\ref{fig:phantom} demonstrates the integration of Tyger within both MaRGE interfaces. Parallel execution was verified, allowing new acquisitions to start while previous datasets are being transmitted and reconstructed without introducing idle scanner time, even under increased transfer times (Table~\ref{tab:latencies}). In Fig.~\ref{fig:phantom}a, distortion correction yields a reconstruction that closely matches the expected geometry, as highlighted by the overlaid contour in red. This confirms the correct implementation of the CP method in compensating for the $B_0$ inhomogeneities of the system, characterized through SPDS-based field mapping integrated in MaRGE. Operationally, this field map can be reused over extended periods under stable conditions. Although CP reconstruction is mathematically related to a Fourier transform, the voxel-dependent off-resonance phase term breaks the separability and coefficient reuse exploited by the FFT (fast Fourier transform), precluding efficient factorization and resulting in a substantially higher computational cost. This burden makes local execution impractical: the same reconstruction required 85\,min when performed offline on the scanner control computer, compared to 21\,s when executed remotely through Tyger (Table~\ref{tab:RARE_params}).

Figure~\ref{fig:knee_den} illustrates the impact of cloud-based denoising on an \emph{in-vivo} knee acquisition, where the improved SNR enables imaging at substantially higher spatial resolution. In this experiment, an in-plane resolution of $0.3 \times 0.3 \times 3.2$\,mm$^3$ was achieved, which is comparable to resolutions routinely used in clinical practice at 3\,T. This result highlights how denoising can be used to push spatial resolution in low-field MRI beyond what is typically feasible when limited by SNR alone. Here, the limiting factor increasingly shifts from SNR to acquisition time, with a total scan duration of 16~min for this T1-w acquisition (Table~\ref{tab:RARE_params}). While classical denoising approaches such as BM4D \cite{Maggioni2013} could be applied, their performance degrades in the low-SNR regime characteristic of LF-MRI, motivating the use of deep-learning-based methods \cite{GuallartNaval2026b}.

Figure~\ref{fig:petra} illustrates the application of cloud-based reconstruction to non-Cartesian PETRA acquisitions, highlighting the ability to process data that cannot be reconstructed locally. Radial reconstructions require iterative methods, like ART, which are computationally prohibitive on the scanner control computer. Using Tyger, the PETRA dataset was successfully reconstructed with ART in approximately 200\,s, whereas equivalent local executions require $\approx20$\,h (Table~\ref{tab:PETRA_params}). Although cloud reconstruction times are still substantial, the parallelized acquisition workflow ensures that they do not extend the overall scan duration. Non-Cartesian sequences of this type are particularly attractive at low field, for example for the visualization of tissues with short $T_2$ components \cite{Borreguero2025b}, further expanding the range of applications enabled by cloud-based processing.

Brain imaging represents a particularly demanding use case, as distortion correction can be relevant due to $B_0$ inhomogeneities that may be further exacerbated for large fields of view. Figure~\ref{fig:brain} shows how distortion correction and denoising are applied jointly in this case, resulting in a substantial improvement in image quality.

\section{Conclusions \& Outlook}

In this work, we have demonstrated the first fully open-source framework that integrates cloud-based processing directly into the acquisition workflow of a low-field MRI scanner. By interfacing MaRCoS/MaRGE with Tyger, we enabled remote and parallel execution of advanced reconstruction pipelines without interrupting ongoing acquisitions. The proposed integration was validated under heterogeneous real-world network conditions and across representative low-field applications.

These results include computationally demanding methods for enhancing the performance of LF-MRI and enabling its routine use, but which are impractical on resource-constrained control computers. In particular, deep-learning denoising enables substantially higher spatial resolution, conjugate-phase distortion correction compensates for the pronounced $B_0$ inhomogeneities inherent to portable Halbach systems, and iterative non-Cartesian reconstructions of radial acquisitions expand the range of feasible contrasts and tissue targets.

Importantly, the applications presented here are merely illustrative examples. The proposed framework is inherently flexible and can incorporate additional reconstruction and processing methods, including established MRI toolboxes such as Gadgetron~\cite{Hansen2013} or BART~\cite{BART}. For example, as denoising shifts the dominant limitation from SNR to acquisition time, acceleration strategies such as compressed sensing are expected to play an increasingly important role \cite{Shimron2025}. The computational demands associated with these approaches further highlight the relevance of scalable cloud-based processing frameworks such as the one presented here.

Looking forward, embedding cloud reconstruction within the acquisition loop opens the door to streaming-based processing, where data could be reconstructed progressively during acquisition and even used to adapt sequence parameters in real time. More broadly, this development provides a foundation for increasingly automated and intelligent LF-MRI systems, in which advanced reconstruction, quantitative analysis, and AI-based interpretation can be deployed independently of local hardware constraints. By decoupling computational capability from scanner cost and location, this framework contributes to making high-quality MRI processing globally accessible.

\appendices

\section*{Contributions}
See Table~\ref{tab:contributions}.

\begin{table*}
\centering
\fontsize{8.5}{9.5}\selectfont
\caption{Author contributions. An “x” indicates participation in the corresponding task.}
\label{tab:contributions}
\begin{tabular}{lccccccccccccccccccc}
\toprule
\textbf{Task} & \textbf{TGN} & \textbf{JS} & \textbf{JMA} & \textbf{HX} & \textbf{JBe} & \textbf{PB} & \textbf{JBo}  & \textbf{JC} & \textbf{FG} & \textbf{PGC} & \textbf{ML} & \textbf{BL} & \textbf{LP} & \textbf{SJS} & \textbf{AW} & \textbf{MH} & \textbf{JA} \\
\midrule
Scanner preparation & x &   & x &   &   & x &   & x & x &   &   &   & x &   &   &   & x  \\
\mg{} coding        & x &   & x &   & x &   &   &   &   &   & x &   &   &   &   &   &    \\
Tyger dev.          &   & x &   & x &   &   &   &   &   &   &   &   &   &   &   & x &    \\
Tyger recons.       & x &   &   & x & x &   &   &   &   &   & x &   &   &   &   &   &    \\
Data acq.           & x &   & x &   &   & x & x &   &   & x &   & x & x &   & x &   & x  \\
Proj. conception    & x &   & x &   &   &   &   &   & x &   &   &   &   &   &   & x & x  \\
Proj. manage.       &   &   &   &   &   &   &   &   &   &   &   &   &   & x &   & x & x  \\
Figure prod.        & x &   &   &   &   &   &   &   &   &   &   &   &   &   &   &   & x  \\
Paper writing       & x &   &   &   &   &   &   &   &   &   &   &   &   &   &   &   & x  \\
Paper revision      & x & x & x & x & x & x & x & x & x & x & x & x & x & x & x & x & x  \\
\bottomrule
\end{tabular}
\normalsize
\end{table*}

\section*{Acknowledgment}
We thank Johnes Obungoloch for allowing us to carry out the data transfer measurements from the Wi-Fi network at MUST.

\section*{Funding}
This work was funded by: the European Innovation Council (NextMRI 101136407), Ministerio de Ciencia e Innovación (PID2022-142719OB-C22), the ISMRM-Gates Knowledge Exchange Program (91484), and US NIH grant 5R01HD085853-12.


\section*{Code availability}
MaRGE (which includes the MaRCoS control system repositories) is available at \url{https://github.com/josalggui/MaRGE}. Instructions for installing Tyger are available at \url{https://microsoft.github.io/tyger/introduction/installation/installation.html}. 

Docker images for distortion correction (from RARE raw data) and non-Cartesian reconstruction (from PETRA raw data) are available at \url{ghcr.io/teresaguallartnaval/dist_corr_tyger:v1} and \url{ghcr.io/teresaguallartnaval/petra_tyger:v2}, respectively. 

All material related to the denoising network (SNRAware) is available at \url{https://github.com/microsoft/SNRAware/}. A Docker image containing the SNRAware repository, together with an interface wrapper that enables its direct execution from within MaRGE, is available at \url{ghcr.io/mrilab-i3m/snraware_local:v1}. 

\section*{Conflict of interest}
TGN consults for PhysioMRI Tech. JMA, FG, and JA are co-founders of PhysioMRI Tech.

\ifCLASSOPTIONcaptionsoff
  \newpage
\fi


\begin{thebibliography}{10}
\providecommand{\url}[1]{#1}
\csname url@samestyle\endcsname
\providecommand{\newblock}{\relax}
\providecommand{\bibinfo}[2]{#2}
\providecommand{\BIBentrySTDinterwordspacing}{\spaceskip=0pt\relax}
\providecommand{\BIBentryALTinterwordstretchfactor}{4}
\providecommand{\BIBentryALTinterwordspacing}{\spaceskip=\fontdimen2\font plus
\BIBentryALTinterwordstretchfactor\fontdimen3\font minus \fontdimen4\font\relax}
\providecommand{\BIBforeignlanguage}[2]{{%
\expandafter\ifx\csname l@#1\endcsname\relax
\typeout{** WARNING: IEEEtran.bst: No hyphenation pattern has been}%
\typeout{** loaded for the language `#1'. Using the pattern for}%
\typeout{** the default language instead.}%
\else
\language=\csname l@#1\endcsname
\fi
#2}}
\providecommand{\BIBdecl}{\relax}
\BIBdecl

\bibitem{McDaniel2019}
\BIBentryALTinterwordspacing
P.~C. McDaniel, C.~Z. Cooley, J.~P. Stockmann, and L.~L. Wald, ``{The MR Cap: A single-sided MRI system designed for potential point-of-care limited field-of-view brain imaging},'' \emph{Magnetic Resonance in Medicine}, vol.~82, no.~5, pp. 1946--1960, nov 2019. [Online]. Available: \url{https://onlinelibrary.wiley.com/doi/full/10.1002/mrm.27861}
\BIBentrySTDinterwordspacing

\bibitem{Nakagomi2019}
M.~Nakagomi, M.~Kajiwara, J.~Matsuzaki, K.~Tanabe, S.~Hoshiai, Y.~Okamoto, and Y.~Terada, ``{Development of a small car-mounted magnetic resonance imaging system for human elbows using a 0.2 T permanent magnet},'' \emph{Journal of Magnetic Resonance}, vol. 304, pp. 1--6, jul 2019.

\bibitem{Cooley2020}
\BIBentryALTinterwordspacing
C.~Z. Cooley, P.~C. McDaniel, J.~P. Stockmann, S.~A. Srinivas, S.~F. Cauley, M.~{\'{S}}liwiak, C.~R. Sappo, C.~F. Vaughn, B.~Guerin, M.~S. Rosen, M.~H. Lev, and L.~L. Wald, ``{A portable scanner for magnetic resonance imaging of the brain},'' \emph{Nature Biomedical Engineering 2020 5:3}, vol.~5, no.~3, pp. 229--239, nov 2020. [Online]. Available: \url{https://www.nature.com/articles/s41551-020-00641-5}
\BIBentrySTDinterwordspacing

\bibitem{OReilly2020}
\BIBentryALTinterwordspacing
T.~O'Reilly, W.~M. Teeuwisse, D.~Gans, K.~Koolstra, and A.~G. Webb, ``{In vivo 3D brain and extremity MRI at 50 mT using a permanent magnet Halbach array},'' \emph{Magnetic Resonance in Medicine}, p. mrm.28396, jul 2020. [Online]. Available: \url{https://onlinelibrary.wiley.com/doi/abs/10.1002/mrm.28396}
\BIBentrySTDinterwordspacing

\bibitem{Sheth2021}
\BIBentryALTinterwordspacing
K.~N. Sheth, M.~H. Mazurek, M.~M. Yuen, B.~A. Cahn, J.~T. Shah, A.~Ward, J.~A. Kim, E.~J. Gilmore, G.~J. Falcone, N.~Petersen, K.~T. Gobeske, F.~Kaddouh, D.~Y. Hwang, J.~Schindler, L.~Sansing, C.~Matouk, J.~Rothberg, G.~Sze, J.~Siner, M.~S. Rosen, S.~Spudich, and W.~T. Kimberly, ``{Assessment of Brain Injury Using Portable, Low-Field Magnetic Resonance Imaging at the Bedside of Critically Ill Patients},'' \emph{JAMA Neurology}, vol.~78, no.~1, pp. 41--47, jan 2021. [Online]. Available: \url{https://jamanetwork.com/journals/jamaneurology/fullarticle/2769858}
\BIBentrySTDinterwordspacing

\bibitem{Mazurek2021}
M.~H. Mazurek, M.~M. Yuen, B.~A. Cahn, M.~S. Rosen, K.~T. Gobeske, E.~J. Gilmore, D.~Hwang, F.~Kaddouh, J.~A. Kim, G.~Falcone, N.~Petersen, J.~Siner, S.~Spudich, G.~Sze, W.~T. Kimberly, and K.~N. Sheth, ``{Low-Field, Portable Magnetic Resonance Imaging at the Bedside to Assess Brain Injury in Patients with Severe COVID-19 (1349)},'' \emph{Neurology}, vol.~96, no. 15 Supplement, 2021.

\bibitem{Liu2021}
\BIBentryALTinterwordspacing
Y.~Liu, A.~T.~L. Leong, Y.~Zhao, L.~Xiao, H.~K.~F. Mak, A.~Chun, O.~Tsang, G.~K.~K. Lau, G.~K.~K. Leung, E.~X. Wu, and X.~Linfang, ``{A low-cost and shielding-free ultra-low-field brain MRI scanner},'' \emph{Nature Communications 2021 12:1}, vol.~12, no.~1, pp. 1--14, dec 2021. [Online]. Available: \url{https://www.nature.com/articles/s41467-021-27317-1}
\BIBentrySTDinterwordspacing

\bibitem{Guallart-Naval2022}
\BIBentryALTinterwordspacing
T.~Guallart-Naval, J.~M. Algar{\'{i}}n, R.~Pellicer-Guridi, F.~Galve, Y.~Vives-Gilabert, R.~Bosch, E.~Pall{\'{a}}s, J.~M. Gonz{\'{a}}lez, J.~P. Rigla, P.~Mart{\'{i}}nez, F.~Lloris, J.~Borreguero, {\'{A}}.~Marcos-Perucho, V.~Negnevitsky, L.~Mart{\'{i}}-Bonmat{\'{i}}, A.~R{\'{i}}os, J.~M. Benlloch, and J.~Alonso, ``{Portable magnetic resonance imaging of patients indoors, outdoors and at home},'' \emph{Scientific Reports 2022 12:1}, vol.~12, no.~1, pp. 1--11, jul 2022. [Online]. Available: \url{https://www.nature.com/articles/s41598-022-17472-w}
\BIBentrySTDinterwordspacing

\bibitem{Obungoloch2023}
J.~Obungoloch, I.~Muhumuza, W.~Teeuwisse, J.~Harper, I.~Etoku, R.~Asiimwe, P.~Tusiime, G.~Gombya, C.~Mugume, M.~H. Namutebi \emph{et~al.}, ``{On-site construction of a point-of-care low-field MRI system in Africa},'' \emph{{NMR in Biomedicine}}, vol.~36, no.~7, p. e4917, 2023.

\bibitem{Guallart-Naval2025c}
\BIBentryALTinterwordspacing
T.~Guallart-Naval, R.~Asiimwe, P.~Tusiime, M.~A. Nassejje, L.~Kinyera, L.~Robin, M.~Nayebare, L.~G.~C. Santos, M.~Fernández-García, L.~Swistunow, J.~M. Algarín, J.~Stairs, M.~Hansen, R.~Amodoi, A.~Webb, J.~Harper, S.~J. Schiff, J.~Obungoloch, and J.~Alonso, ``In-vivo imaging with a low-cost mri scanner and cloud data processing in low-resource settings,'' 2025. [Online]. Available: \url{https://arxiv.org/abs/2511.19226}
\BIBentrySTDinterwordspacing

\bibitem{Sarracanie2020}
\BIBentryALTinterwordspacing
M.~Sarracanie and N.~Salameh, ``{Low-Field MRI: How Low Can We Go? A Fresh View on an Old Debate},'' \emph{Frontiers in Physics}, vol.~8, p. 172, jun 2020. [Online]. Available: \url{https://www.frontiersin.org/article/10.3389/fphy.2020.00172/full}
\BIBentrySTDinterwordspacing

\bibitem{Webb2023b}
A.~Webb and T.~O’Reilly, ``{Tackling SNR at low-field: a review of hardware approaches for point-of-care systems},'' \emph{Magnetic Resonance Materials in Physics, Biology and Medicine}, vol.~36, no.~3, pp. 375--393, 2023.

\bibitem{Borreguero2025b}
\BIBentryALTinterwordspacing
J.~Borreguero, L.~G.~C. Santos, L.~V. Cid, E.~Castañón, M.~Fernández-García, P.~Benlloch, R.~Bosch, J.~Conejero, P.~García-Cristóbal, A.~González-Cebrián, T.~Guallart-Naval, E.~Pallás, L.~Porcar, L.~Swistunow, J.~M. Algarín, F.~Galve, and J.~Alonso, ``{Qualitative and quantitative hard-tissue MRI with portable Halbach scanners},'' \emph{arXiv preprint arXiv:2511.15617}, 2025. [Online]. Available: \url{https://arxiv.org/abs/2511.15617}
\BIBentrySTDinterwordspacing

\bibitem{Zhao2024}
\BIBentryALTinterwordspacing
Y.~Zhao, Y.~Ding, V.~Lau, C.~Man, S.~Su, L.~Xiao, A.~T.~L. Leong, and E.~X. Wu, ``{Whole-body magnetic resonance imaging at 0.05 tesla},'' \emph{Science}, vol. 384, 5 2024. [Online]. Available: \url{https://www.science.org/doi/10.1126/science.adm7168}
\BIBentrySTDinterwordspacing

\bibitem{Islam2023}
K.~T. Islam, S.~Zhong, P.~Zakavi, Z.~Chen, H.~Kavnoudias, S.~Farquharson, G.~Durbridge, M.~Barth, K.~L. McMahon, P.~M. Parizel, A.~Dwyer, G.~F. Egan, M.~Law, and Z.~Chen, ``{Improving portable low-field MRI image quality through image-to-image translation using paired low- and high-field images},'' \emph{Scientific Reports}, vol.~13, p. 21183, 2023.

\bibitem{Shimron2025}
\BIBentryALTinterwordspacing
E.~Shimron, L.~Shapira, D.~Raviv, and O.~Cohen, ``{Accelerating Low-field MRI: From Compressed Sensing to Deep Learning Reconstruction with CNNs and Transformers},'' \emph{arXiv preprint arXiv:2411.06704}, 2025. [Online]. Available: \url{https://arxiv.org/abs/2411.06704}
\BIBentrySTDinterwordspacing

\bibitem{Koolstra2021}
K.~Koolstra, T.~O'Reilly, P.~Börnert, and A.~Webb, ``{Image distortion correction for MRI in low field permanent magnet systems with strong B0 inhomogeneity and gradient field nonlinearities},'' \emph{Magnetic Resonance Materials in Physics, Biology and Medicine (MAGMA)}, vol.~34, no.~5, pp. 631--642, 2021.

\bibitem{Borreguero2025}
\BIBentryALTinterwordspacing
J.~Borreguero, F.~Galve, J.~M. Algarín, and J.~Alonso, ``Zero-echo-time sequences in highly inhomogeneous fields,'' \emph{Magnetic Resonance in Medicine}, vol.~93, no.~3, pp. 1190--1204, 2025. [Online]. Available: \url{https://onlinelibrary.wiley.com/doi/abs/10.1002/mrm.30352}
\BIBentrySTDinterwordspacing

\bibitem{Schote2025}
D.~J. Schote, I.~Bhattacharya, J.~Rieger, X.~Golay, T.~O'Reilly, and A.~Webb, ``{Nexus: A versatile console for advanced low-field MRI},'' \emph{Magnetic Resonance in Medicine}, 2025.

\bibitem{Negnevitsky2023}
\BIBentryALTinterwordspacing
V.~Negnevitsky, Y.~Vives-Gilabert, J.~M. Algar\'in, L.~Craven-Brightman, R.~Pellicer-Guridi, T.~O'Reilly, J.~P. Stockmann, A.~Webb, J.~Alonso, and B.~Menk\"uc, ``{MaRCoS, an open-source electronic control system for low-field MRI},'' \emph{Journal of Magnetic Resonance}, vol. 350, p. 107424, 2023. [Online]. Available: \url{https://doi.org/10.1016/j.jmr.2023.107424}
\BIBentrySTDinterwordspacing

\bibitem{Guallart-Naval2022b}
\BIBentryALTinterwordspacing
T.~Guallart-Naval, T.~O'Reilly, J.~M. Algarín, R.~Pellicer-Guridi, Y.~Vives-Gilabert, L.~Craven-Brightman, V.~Negnevitsky, B.~Menküc, F.~Galve, J.~P. Stockmann, A.~Webb, and J.~Alonso, ``{Benchmarking the performance of a low-cost magnetic resonance control system at multiple sites in the open MaRCoS community},'' \emph{NMR in Biomedicine}, vol.~36, no.~1, p. e4825, 2023. [Online]. Available: \url{https://analyticalsciencejournals.onlinelibrary.wiley.com/doi/abs/10.1002/nbm.4825}
\BIBentrySTDinterwordspacing

\bibitem{Algarin2024}
{Algar{\'\i}n, Jos{\'e} M and Guallart-Naval, Teresa and Borreguero, Jos{\'e} and Galve, Fernando and Alonso, Joseba}, ``{MaRGE: A graphical environment for MaRCoS},'' \emph{{Journal of Magnetic Resonance}}, vol. 361, p. 107662, 2024.

\bibitem{Tyger}
\BIBentryALTinterwordspacing
``Github - microsoft/tyger: Remote signal processing.'' [Online]. Available: \url{https://github.com/microsoft/tyger}
\BIBentrySTDinterwordspacing

\bibitem{Galve2024}
\BIBentryALTinterwordspacing
F.~Galve, E.~Pallás, T.~Guallart-Naval, P.~García-Cristóbal, P.~Martínez, J.~M. Algarín, J.~Borreguero, R.~Bosch, F.~Juan-Lloris, J.~M. Benlloch, and J.~Alonso, ``Elliptical halbach magnet and gradient modules for low-field portable magnetic resonance imaging,'' \emph{NMR in Biomedicine}, vol.~37, no.~12, p. e5258, 2024. [Online]. Available: \url{https://analyticalsciencejournals.onlinelibrary.wiley.com/doi/abs/10.1002/nbm.5258}
\BIBentrySTDinterwordspacing

\bibitem{Ravi2019}
\BIBentryALTinterwordspacing
K.~S. Ravi, S.~Geethanath, and J.~T. Vaughan, ``{PyPulseq: A Python Package for MRI Pulse Sequence Design},'' \emph{Journal of Open Source Software}, vol.~4, no.~42, p. 1725, 2019. [Online]. Available: \url{https://doi.org/10.21105/joss.01725}
\BIBentrySTDinterwordspacing

\bibitem{Hansen2013}
M.~S. Hansen and T.~S. Sørensen, ``{Gadgetron: An open source framework for medical image reconstruction},'' \emph{Magnetic Resonance in Medicine}, vol.~69, no.~6, pp. 1768--1776, 2013.

\bibitem{BART}
M.~Uecker, M.~Lustig \emph{et~al.}, ``{BART: Berkeley Advanced Reconstruction Toolbox},'' \url{https://github.com/mrirecon/bart}.

\bibitem{GuallartNaval2026a}
\BIBentryALTinterwordspacing
T.~Guallart-Naval, J.~M. Algarín, and J.~Alonso, ``{Electromagnetic Noise Characterization and Suppression in Low-Field MRI Systems},'' \emph{{Magnetic Resonance in Medicine}}, vol.~95, no.~5, pp. 3000--3007. [Online]. Available: \url{https://onlinelibrary.wiley.com/doi/abs/10.1002/mrm.70235}
\BIBentrySTDinterwordspacing

\bibitem{MRD}
\BIBentryALTinterwordspacing
``{ISMRMRD / MRD: Magnetic Resonance Data format}.'' [Online]. Available: \url{https://ismrmrd.github.io/mrd/}
\BIBentrySTDinterwordspacing

\bibitem{Xue2025}
\BIBentryALTinterwordspacing
H.~Xue, S.~M. Hooper, I.~Pierce, R.~H. Davies, J.~Stairs, J.~Naegele, A.~E. Campbell-Washburn, C.~Manisty, J.~C. Moon, T.~A. Treibel, M.~S. Hansen, and P.~Kellman, ``{SNRAware: Improved Deep Learning MRI Denoising with Signal-to-Noise Ratio Unit Training and G-Factor Map Augmentation},'' \emph{Radiology: Artificial Intelligence}, vol.~7, no.~6, p. e250227, 2025, pMID: 41123451. [Online]. Available: \url{https://doi.org/10.1148/ryai.250227}
\BIBentrySTDinterwordspacing

\bibitem{Hennig1986}
J.~Hennig, A.~Nauerth, and H.~Friedburg, ``{RARE imaging: A fast imaging method for clinical MR},'' \emph{Magnetic Resonance in Medicine}, vol.~3, no.~6, pp. 823--833, 1986.

\bibitem{Kellman2005}
P.~Kellman and E.~R. McVeigh, ``{Image reconstruction in SNR units: A general method for SNR measurement},'' \emph{Magnetic Resonance in Medicine}, vol.~54, no.~6, pp. 1439--1447, 2005.

\bibitem{Noll1991}
D.~C. Noll, C.~H. Meyer, J.~M. Pauly, D.~G. Nishimura, and A.~Macovski, ``A homogeneity correction method for magnetic resonance imaging with time-varying gradients,'' \emph{IEEE transactions on medical imaging}, vol.~10, no.~4, pp. 629--637, 1991.

\bibitem{Noll2005}
D.~C. Noll, J.~A. Fessler, and B.~P. Sutton, ``Conjugate phase mri reconstruction with spatially variant sample density correction,'' \emph{IEEE transactions on medical imaging}, vol.~24, no.~3, pp. 325--336, 2005.

\bibitem{Grodzki2012}
D.~M. Grodzki, P.~M. Jakob, and B.~Heismann, ``{Ultrashort echo time imaging using pointwise encoding time reduction with radial acquisition (PETRA)},'' \emph{Magnetic Resonance in Medicine}, vol.~67, no.~2, pp. 510--518, feb 2012.

\bibitem{Gordon1970}
R.~Gordon, R.~Bender, and G.~T. Herman, ``{Algebraic Reconstruction Techniques (ART) for three-dimensional electron microscopy and X-ray photography},'' \emph{Journal of Theoretical Biology}, vol.~29, no.~3, pp. 471--481, dec 1970.

\bibitem{Algarin2020}
\BIBentryALTinterwordspacing
J.~M. Algar{\'{i}}n, E.~D{\'{i}}az-Caballero, J.~Borreguero, F.~Galve, D.~Grau-Ruiz, J.~P. Rigla, R.~Bosch, J.~M. Gonz{\'{a}}lez, E.~Pall{\'{a}}s, M.~Corber{\'{a}}n, C.~Gramage, S.~Aja-Fern{\'{a}}ndez, A.~R{\'{i}}os, J.~M. Benlloch, and J.~Alonso, ``{Simultaneous imaging of hard and soft biological tissues in a low-field dental MRI scanner},'' \emph{Scientific Reports}, vol.~10, no.~1, p. 21470, 2020. [Online]. Available: \url{https://doi.org/10.1038/s41598-020-78456-2}
\BIBentrySTDinterwordspacing

\bibitem{ACRphantom}
{American College of Radiology}, \emph{{MRI Quality Control Manual}}, American College of Radiology, Reston, VA, USA, 2015.

\bibitem{Maggioni2013}
M.~Maggioni, V.~Katkovnik, K.~Egiazarian, and A.~Foi, ``{Nonlocal transform-domain filter for volumetric data denoising and reconstruction},'' \emph{IEEE Transactions on Image Processing}, vol.~22, no.~1, pp. 119--133, 2013.

\bibitem{GuallartNaval2026b}
T.~Guallart-Naval, H.~Xue, and \emph{et al.}, ``{AI denoising with SNRAware in low-field MRI},'' \emph{In preparation}, 2026.

\end{thebibliography}

\end{document}